**Self-induced crystalline fluctuation spin-glass state in $Mn_7C_3$ binary compounds**

Zekun Yu[1,2], Chao Zhou[1], Kuo Bao[1*], Xiaofeng Wang[3], Zhaoqing Wang[1], Jinming Zhu[1], Enxuan Li[1], Andong Yao[1], Yuhan Meng[1], Yufei Ge[1], Xingbin Zhao[4], Shuailing Ma[4], Pinwen Zhu[1], Qiang Tao[1], Tian Cui[1,4*]

[1]State Key Laboratory of Superhard Materials, College of Physics, Jilin University, Changchun, 130012, China

[2]Center for Optics Research and Engineering, Shandong University, Qingdao, 266237, China

[3]State Key Laboratory of Inorganic Synthesis and Preparative Chemistry, College of Chemistry, Jilin University, Chang chun 130012, China

[4]Institute of High Pressure Physics, School of Physical Science and Technology, Ningbo University, Ningbo 315211, China

*Corresponding authors: baokuo@jlu.edu.cn, and cuitian@nbu.edu.cn

**Abstract**

Crystalline spin glasses are attractive compounds owing to their unique nature and applications. Here, we synthesised a bulk *Pnma*-type $Mn_7C_3$ spin glass by a high-temperature, high-pressure method. Experimental characterisation including X-ray diffraction and magnetic susceptibility measurements demonstrated that the compound has a triangular Ising-model-based structure, high freezing temperature of 37.4 K, and novel competition mechanism. Theoretical calculations and simulations revealed that the triangular Mn units are spontaneously frustrated and bridge neighbouring Mn units via polarised C atoms and messenger Mn atoms. Triangular C units each share one electron within a three-pronged electron cloud. This electron is the direct cause of frustration and competition in $Mn_7C_3$. The competition within the triangular Mn units suggests that the possible magnetic configurations are highly degenerate and that the $Mn_7C_3$ spin glass has high robustness. This work introduces a new family of spin glasses with ordered microgeometries that drive electronic structure disorder, and an application-friendly spin-glass material for use in fields like high-efficiency hardware and algorithm design in artificial intelligence.

## Introduction

Owing to their unique triangular Ising-model-based frustration structures that mimic neuronal interactions, spin glasses are promising materials for important applications including artificial intelligence[1], neural networks[2, 3], and deep learning[4]. In addition, they are believed to play a crucial role in iron-based superconductors[5] and quantum topological excitations[6–9]. However, a comprehensive understanding of the properties of these materials is lacking. Unlike ferromagnetic and antiferromagnetic materials, spin glasses exhibit no long-range spin order even at low temperatures, yet their spins are ordered in a temporal dimension. Thus, spin glasses are considered `contradictory' systems as they are with and without spin order at the same time[10].

In accordance with their name, spin glasses usually exhibit some glassy microstructural characteristics, such as atomic disorder or random replacement. Some examples of spin glasses include dilute magnetic alloys such as $C_{1-x}Mn_x$[11], site-disordered crystals such as $Fe_2TiO_5$[12], and random atomic displacement crystals such as $Y_2Mo_2O_7$[13]. Well-ordered spin glasses exhibit more robustness than disordered ones; furthermore, the ordered structure can be easily manipulated and conveniently modelled. Recently several types of well-ordered spin glasses have been discovered, including kagome spin glasses[14] and elementary spin glasses[6]. Saccone *et al.[15]* realised an artificial spin glass based on the principles of Hopfield neural networks. The triangular Ising theory gives a good indication of this structure, wherein three spins are evenly spaced at the vertices of a triangle with frustrated antiferromagnetic Ising interactions in two dimensions[16].

Therefore, well-ordered compounds with hexagonal or triangular structures with magnetic elements at the vertices of the lattice are ideal candidates for spin glasses. However, well-ordered spin glasses still have many problems that limit their further development and application, such as an inability to be mass produced, a low density of triangular units, and a low freezing temperature. Herein, we fond and present a new kind of natural well-ordered spin glass, $Mn_7C_3$, which exhibits a novel competition and correlation mechanism within its frustration structure and put forward a new spin exchange-correlation model to explain this phenomenon. The Mn atoms have varying valence states with different magnetic moments. In addition, the C atoms, which have a suitable size for use as interval atoms between the Mn atoms, have several valence states. Furthermore, the pressure effect enables tuning of the exchange between the Mn atoms. Compared with other spin glasses, this novel material has a high density of triangular Ising units, a higher freezing temperature, and lower cost. Moreover, its high-pressure synthesis method is environmentally friendly and suited to mass production.

Therefore, spin glasses comprising $Mn_7C_3$ with well-ordered lattices can be widely used in hardware and algorithm design for artificial intelligence applications and provide a new material space for the exploration and design of other spin-glass materials. It is noteworthy that this discovery opens the door to a new family of spin glasses whose ordered microgeometries of matter. But it will also provide a proving ground to develop new theories where we can link physics to other fields, for example, theoretical neuroscience mentioned earlier.

**Experimental methods and details**

**Preparation and synthesis:** We synthesised polycrystalline bulk $Mn_7C_3$ by using a high-temperature and high-pressure method in a six-anvil press (SPT 6 × 14400 N). The sample was pre-processed by mixing Mn (Alpha *Co., Ltd.*, CAS: 7439-96-5; APS 50-100 μm, purity 99.6%) and C (Aladdin *Co., Ltd.*, CAS: 7782-42-5; APS 1-5 μm, purity 99.95%) powders at the nominal molar ratio in an agate mortar (φ 50 mm) for 4 h. We then pre-compressed the mixed powder into a cylinder with a diameter and height of 4.0 and 2.5 mm, respectively. Thereafter, we placed the cylinder in a specific cell (*SI Appendix*, Fig. S1) to prevent contact with moisture and oxygen from the atmosphere, and carried out the synthesis. The temperature and pressure for synthesis were 2300 K and 5.0 GPa, respectively. The sample was held under these conditions for 30 min and then quenched to room temperature at an approximate cooling rate of 200 K/s.

**Experiment and Characterisation:** X-ray diffraction was conducted using the XtaLAB Synergy Custom system (Rigaku Corp. *Inc.*)[26]. $Mn_7C_3$ emitted strong secondary radiation when using a Cu target, so spectra were acquired using a Mo target. The sample was tested by using a high-flux rotating anode X-ray source (FR-X) with a power of 2.97 kW and integration time of 60 s at room temperature and pressure. Data refinement was conducted using the Rietveld method and Le Bail method in the General Structure Analysis System (GSAS) software package[27]. The crystal structure parameters were obtained from the Inorganic Crystal Structure Database (ICSD) with collection code 31017. Structural and morphological characterisations were carried out using the Magellan 400 (XHR) SEM (Field Electron and Ion *Co.*), with EDS microanalysis using an Ultim Max Silicon Drift Detector and AZtec system (Oxford Instruments *Plc.*). The acquisition parameters were as follows: spot size, 3 nm; accelerating voltage, 3.00 kV; current, 25 pA; working distance, 3.8 mm; and detector mode, ETD. The DC/AC magnetic susceptibility was measured using a Physical Property Measurement System and Superconducting Quantum Interference Device-Vibrating Sample Magnetometer (Quantum Design *Co., Ltd.*)[28, 29]. All susceptibilities were measured using a 10 Oe field in the temperature range of 5-300 K. Fitting of the AC susceptibility was conducted by using the slowing down model[30].

Electrochemical impedance spectroscopy was conducted using the SI-1296 electrochemical interface and SI-1260 interface/grain phase analyser (Solartron-Mobrey *Inc.*). The AC amplitude was 10 mV and without DC potential. The fitting results were acquired using ZView software[31].

**Theoretical Calculation and Simulation:** The structural relaxation and properties of the material were calculated with a density functional theory (DFT)[32] framework and the projector augmented-wave (PAW)[33] method using the Vienna Ab-initio simulation package (VASP)[34]. The initial crystal structure file was selected based on our XRD refinement results (*SI Appendix*, Table S1). The Perdew-Burke-Ernzerhof generalised gradient approximation (PBE-GGA)[35] was used to deal with the exchange-correlation potential. An energy cut-off of 600 eV for the plane-wave basis and Monkhorst-Pack k-mesh spacing of $2\pi\times0.03$ Å$^{-1}$ were used in the calculations. The forces between the atoms converged to within 0.01 eV/Å in the relaxation. Electron configurations of $3p^6 3d^6 4s^1$ and $1s^2 2p^1$ were taken as the valence states of Mn and C, respectively. To calculate the magnetic moments, several representative magnetic configurations were considered. The parallel state used the wavefunction $|\chi_1\rangle_{i,i+1}=|\psi\rangle_i+|\psi\rangle_{i+1}$, $i$=1,3⋯13, and the antiparallel state used the wavefunction $|\chi_2\rangle_{i,i+1}=|\psi\rangle_i+|\varphi\rangle_{i+1}$, $i$=1,3, ⋯,13. The wavefunction of the random state was randomly placed by using the crypto.getRandomValues algorithm.

**Others:** All charts and contours were drawn using Origin and VESTA[36] software, and all crystal structures were modelled and rendered in Unigraphics NX and PhotoView 360 software, respectively.

## Results and discussion

We synthesised polycrystalline bulk $Mn_7C_3$ by a high-temperature, high-pressure method. Scanning electron microscopy (SEM) and related analyses (see *SI Appendix*, Table S1 and *SI Appendix*, Fig. S2) revealed that the sample has good crystallinity and almost no pores or chasms. The X-ray diffraction pattern (Fig. 1A) and Rietveld refinement data (*SI Appendix*, Table S2) confirm that the material is crystalline with the *Pnma* space group. The crystal contains five inequivalent Mn sites and two equivalent C sites (Fig. 1B). Combined with our calculation results, there are two main units in the $Mn_7C_3$ microstructure, namely, an outer honeycomb hexagonal structure (Fig. 1C), which comprises vertex and zig-zag Mn atoms (referred to as `messenger' Mn atoms), and two medial Mn triangles of `frustrated' Mn atoms (Fig. 1D) with adjacent C triangles. The reasoning behind these names is discussed later.

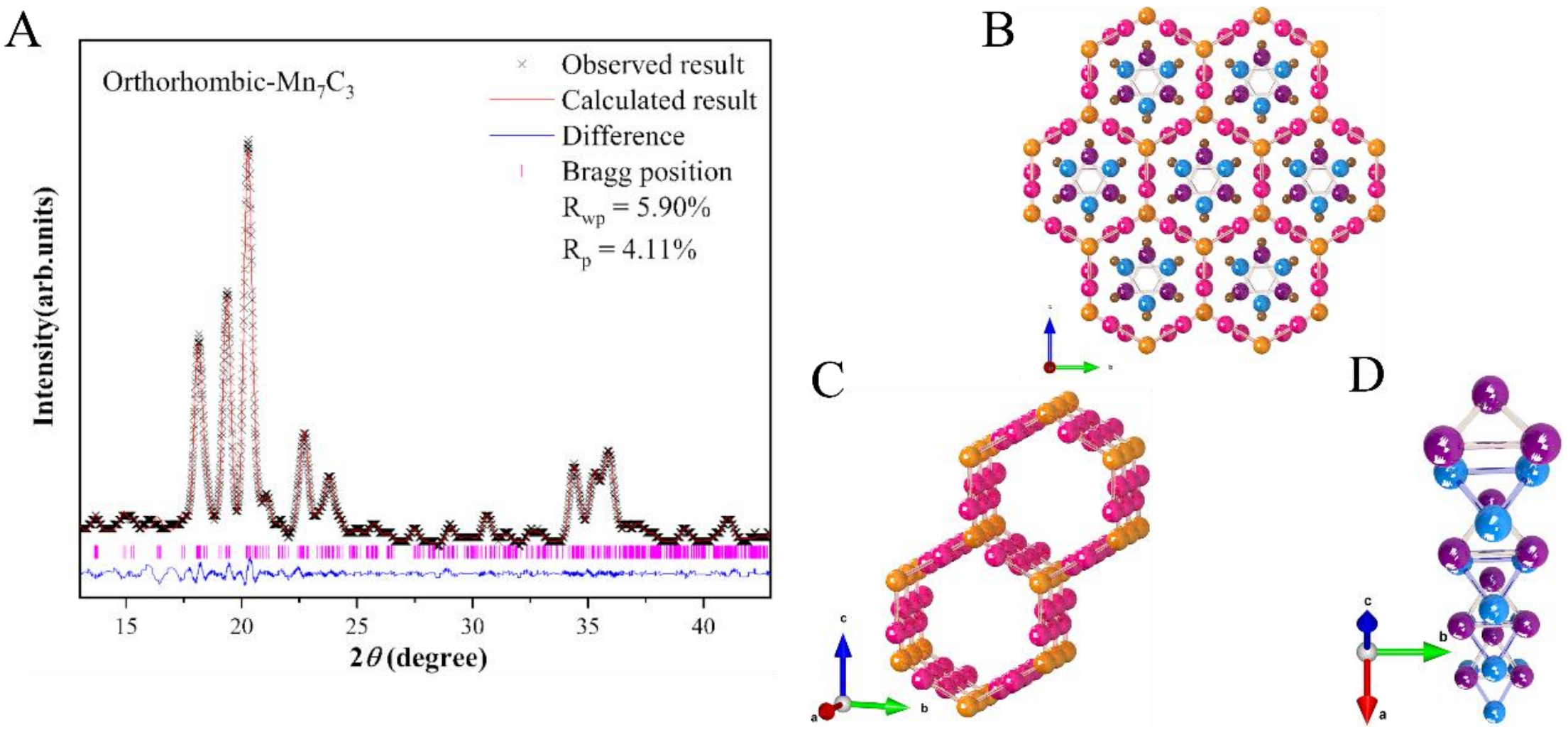

**Fig. 1.** Structure and typical micro-sub-units of $Mn_7C_3$. (A) Rietveld refinement pattern from X-ray diffraction data; (B) Crystalline structure along the a-axis of $Mn_7C_3$; (C) Lateral Mn hexagonal units; (D) Medial Mn triangular units. The pink and orange balls are zig-zag and vertex Mn atoms, respectively, in the honeycomb hexagonal units; the blue and purple balls are Mn atoms in the medial triangular units, where different colours refer to different layers; and the small brown balls are the C atoms adjacent to the Mn atoms in the triangular units.

The magnetic susceptibility curves of $Mn_7C_3$ with zero-field cooling (ZFC) and field cooling (FC) contained lambda peaks that are typical of spin-glass materials (Fig. 2A). By calculating the second derivatives of the curves (*SI Appendix*, Fig. S3A), we found that the spins start to freeze at 42 K, with total freezing at 37.4 K. When the compound was cooled further, both curves suddenly increased at 25 K instead of reaching a limit. This phenomenon is explained below. When we studied the real part of the AC magnetic susceptibility at different frequencies $f$ (Fig. 2B), a broad peak appeared at 37.4 K when $f = 1$ Hz, which decreased in intensity and shifted to higher temperatures as $f$ increased. The imaginary parts of all frequencies were zero across the entire temperature range, except at 25-42 K (*SI Appendix*, Fig. S3B). This indicates that the correlation lengths of the interacting spins within $Mn_7C_3$ change gradually at 25-42 K[17], and that there is no connection between the spin-glass behaviour and the magnetic susceptibility, except for the freezing process. Meanwhile, our data meet the universal formula of the slowing down model[10, 18],

$$\tau = \tau_0 \left[\frac{T_f - T_{SG}}{T_{SG}}\right]^{-z_\nu}$$

where $\tau$ and $\tau_0$ are the characteristic time constant and relaxation time of the spin-glass system, respectively; $T_{SG}$ and $T_f$ are the freezing temperatures at 0 Hz and the test frequency f, respectively; and $z_v$ is a dynamic index that is dependent on the system. Normally, the value of $z_v$ is between 4 and 12 according to the family of spin glass, and increases with an increase in conductivity. It typically decreases in the following order: transition-metal solutes > rare-earth combinations > amorphous (metallic) glasses > dilute magnetic semiconductors > insulating spin glasses. Good fitting of the equation with $T_{SG}$ = 37.4 K and $\tau_0$= 1.69 × $10^{12}$ s was achieved when $z_v$ = 5, as shown in Fig. 2C. This value is closest to that of transition-metal solute spin glasses such as $Cu_{1-x}Mn_x$ and $Au_{1-x}Fe_x$[19–21]; however, the derivation of the fitting line is closer to that of dilute magnetic semiconductor spin glasses such as $YbMgGaO_4$[22]. The underlying physics for this phenomenon is discussed in detail later. Furthermore, we measured the AC susceptibility with different DC background fields, and found that a weak field of a few hundred Gauss could prohibit spin freezing. This is characteristic of natural spin glasses owing to their poor robustness[23] (see Fig. 2D). There were no observable peaks in the curves when the background field was above 1000 Oe. Meanwhile, the curves were nearly identical to those in Fig. 2A below 25 K, regardless of the applied DC field. Notably, unlike any known type of spin glass, $Mn_7C_3$ appears to be a new kind of bulk crystalline spin glass that relies on a specific crystal structure instead of disorder to achieve spin frustration.

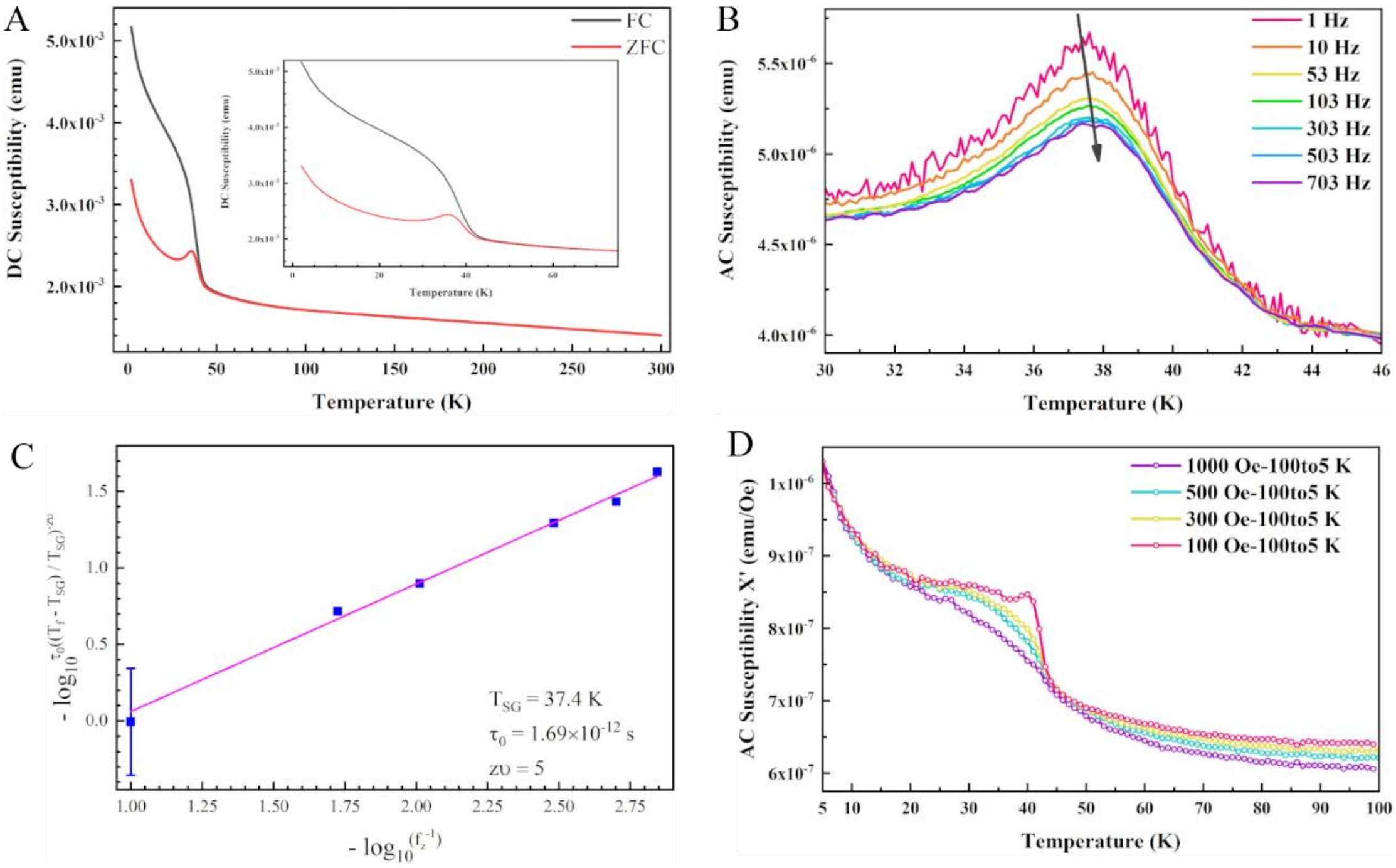

**Fig. 2.** Magnetic susceptibility of the $Mn_7C_3$ sample. (A) DC magnetic susceptibility $\chi$ with field cooling (FC) (black line) and zero-field cooling (ZFC) (red line). The inset figure is an enlarged view of the low temperature region of the graph. (B) Temperature dependence of the real part of the AC magnetic susceptibility $\chi'$ at AC field frequencies of 1 to 703 Hz. (C) Frequency dependence of the freezing temperature fitted using the slowing down theoretical mode. (D) AC magnetic susceptibility with DC backgrounds of 100 to 1000 Oe.

To acquire a deep understanding of this unprecedented material, we simulated its magnetic and electronic structure. $Mn_7C_3$ is a complex system for *ab intio* simulations because there are 28 Mn and 12 C atoms within its smallest simulation cell. If all the possible magnetic states were to be considered, its simulation would be even more challenging. Without considering spin polarisation, $Mn_7C_3$ is stable and its symmetry remains in the *Pnma* (No. 62) space group. However, if its magnetic states are considered, the spins break the time inversion symmetry, and the symmetry of the system breaks to the *P*-1 (No. 2) space group. We labelled the two Mn atoms with inverse symmetry, whereby the *i*th Mn atom correlates with the inverse -*i*th one, and the wavefunctions of the pair are[24]

$$|\psi\rangle_i = |m,\uparrow\rangle_i + |m,\downarrow\rangle_{-i}$$

$$|\phi\rangle_i = |m,\downarrow\rangle_i + |m,\uparrow\rangle_{-i}$$

where *m* is the absolute moment of the *i*th Mn atom. If only spin direction is considered, there are $2^{14}$ possible magnetic configurations for $Mn_7C_3$; however, if the absolute moment of each Mn atom were to be considered, the total number of configurations might be infinite. This is clearly too many magnetic configurations to realistically study. Therefore, we carefully selected several representative magnetic configurations, namely, random, parallel, antiparallel with zero moment, and antiparallel with non-zero moment. The detailed configurations of these states are listed in (*SI Appendix*, Table. S3). After full conversion, the relaxation energies of these configurations were -365.86, -365.91, -365.89, and -365.86 eV, respectively. The absolute magnetic moment for each Mn atom in these configurations was distributed from 0 to 1.6 $\mu_B$.

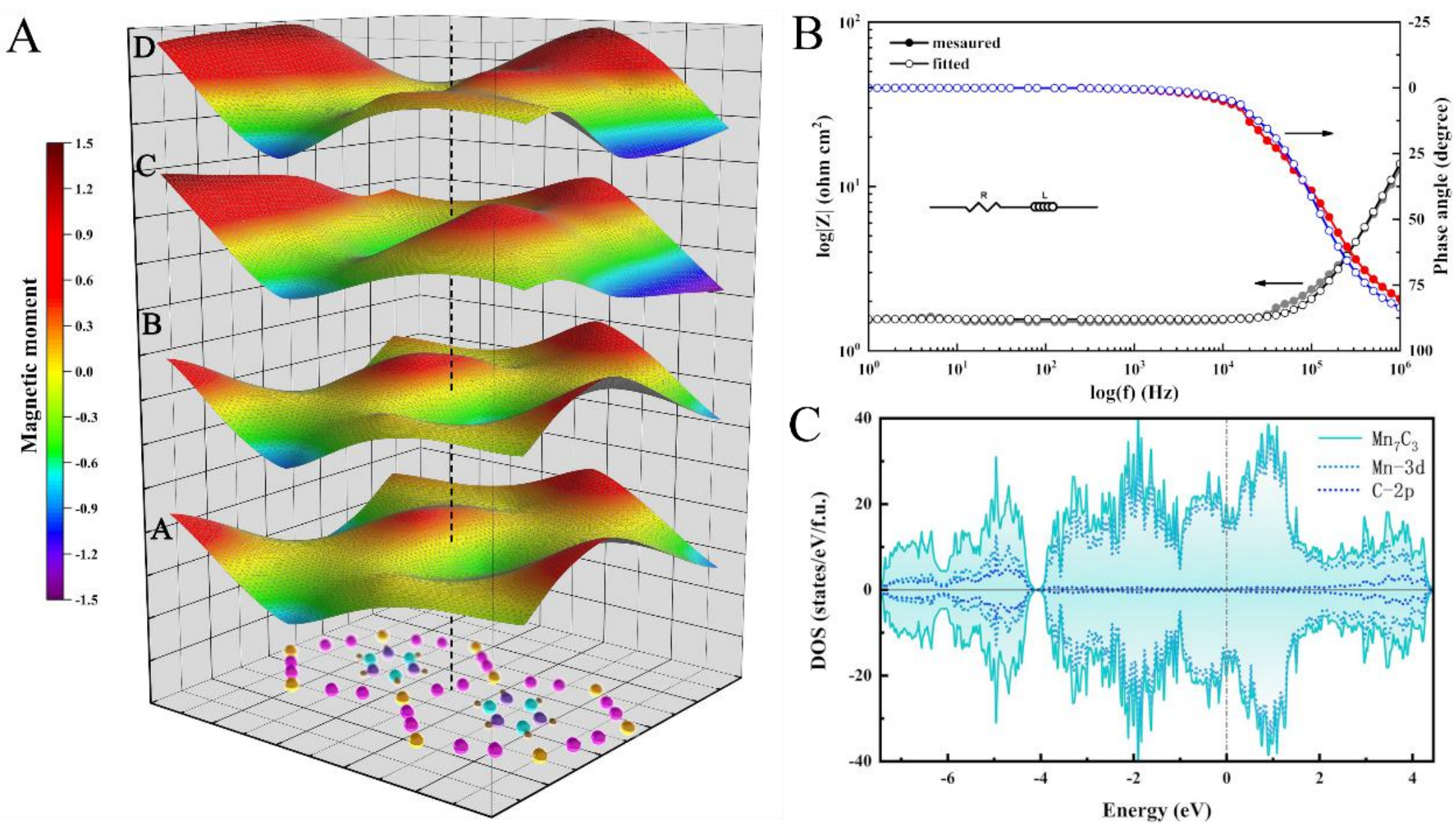

**Fig. 3.** Magnetic distribution and electronic properties of $Mn_7C_3$. (A) Magnetic distribution maps of $Mn_7C_3$ with different initial settings. A: random state, B: parallel state, C: antiparallel state with zero moment, and D: antiparallel state with non-zero moment. The positions of the atoms within $Mn_7C_3$ are projected on the x-y plane along the a-axis. The z-axis height and colour coding represent the moment of each atom, and the vertical dotted line indicates the axis of symmetry. (B) Impedance spectra of $Mn_7C_3$. The solid and hollow circles represent the measured and fitted results, respectively. The inset image shows the equivalent fitting circuit. (C) Electronic density of states (DOS) of $Mn_7C_3$. The plots along the positive and negative y-axes represent the spin-up and spin-down states, respectively.

We designed a three-dimensional (3D) map to illustrate the above results (Fig. 3A). The colours and peaks indicate the relative strength of the magnetic field of the $Mn_7C_3$ crystal cell in different areas. With careful study, we found commonalities for all the possible configurations; if only the spin directions are considered, all spin configurations have a symmetry of *P*-1 around the origin. Furthermore, regarding the converged energies with different magnetic configurations, the difference between them is small enough to claim that they are highly degenerate, and that minor thermal differences might alter them from one magnetic configuration to another. This proves that there might be infinite degenerate magnetic phases of $Mn_7C_3$; that is, it must be a disordered magnetic system with frustrated magnetism, for which it is impossible to form any magnetic order in a certain direction. The AC/DC susceptibility measurement in Fig. 2D also supports this result.

In addition, there is a spin exchange path in the $Mn_7C_3$ system, which we address later based on the electronic structure. The partial density of states (PDOS) in Fig. 3C indicates that $Mn_7C_3$ is a conductor, with the Mn 3*d* and C 2*p* electrons strongly hybridised in the range of -7.5 to -3.0 eV. We measured its conductivity by electronic impedance spectroscopy (EIS) with a RL circuit description code. The EIS results and a schematic of the equivalent circuit are shown in Fig. 3B. The zero phase angle at low frequency means that the imaginary term is zero; thus, the impedance Z is equivalent to the conductivity R. The resistivity of the fitting result is above $2.81\times10^{-3}$ Ω·m, which is similar to that of traditional dilute magnetic alloys as $z_v$. A marginal inductive reactance signal is present at 300 kHz, which is sign of the eddy current of Mn 3*d* electrons. This eddy current is produced by the skin effect and affects the signal quality of the imaginary part of AC susceptibility. Moreover, we found that the conductivity of $Mn_7C_3$ was very sensitive to the DC magnetic field, similarly to the magnetic susceptibility results in Fig. 2D. This phenomenon will be explored and discussed further in another paper.

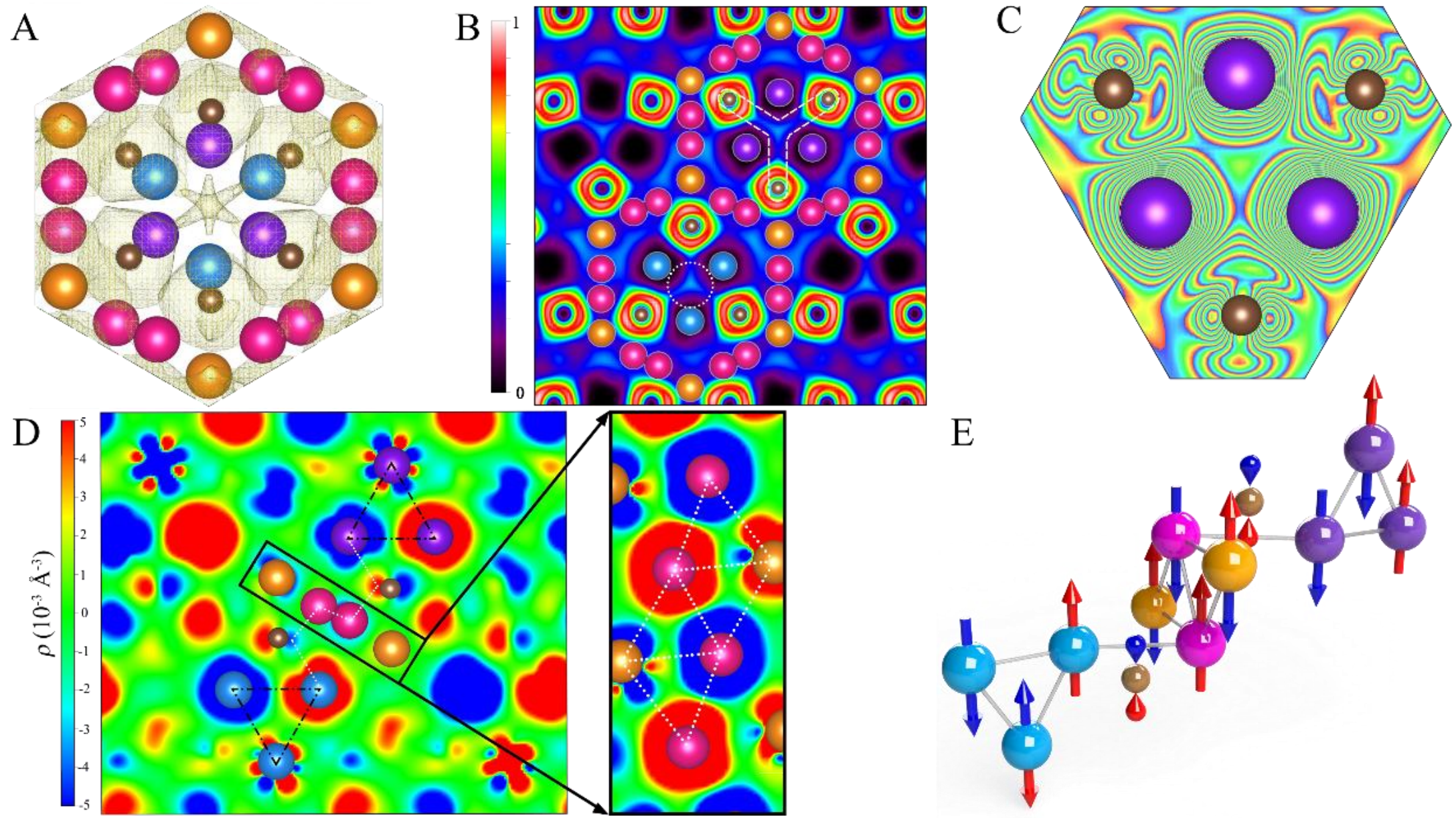

**Fig. 4.** Electron localisation function maps and spin frustration model of $Mn_7C_3$. (A) 3D map of a hexagonal unit with a 0.16 eV iso-surface. (B) Map of localised electrons in two $Mn_7C_3$ structures. The three-pronged dashed line and circular dotted line indicate the triangular electronic passage of C atoms and its centre, respectively. (C) Electron localisation function of a medial triangular unit in the (100) plane. (D) Spin density of $Mn_7C_3$. The white dotted lines represent exchange-correlation effects between atoms. From top to bottom, the zones marked by three black dot-dashed lines are the triangular structures. One of the chains of a honeycomb hexagonal structure and another triangular structure form the neighbouring honeycomb hexagonal structure. The square zone is the one side of hexagonal structure with zig-zagged atoms and vertex atoms. The right panel shows an enlarged view of the area in the black rectangle. The white dotted lines

represent exchange-correlation effects between atoms. (E) Spin frustration model of $Mn_7C_3$. Different size and colour spheres represent different atoms in $Mn_7C_3$, as described in the caption of Fig. 1. The small red and blue cones represent the spin polarisation of C atoms, and the red and blue arrows represent the spin direction.

Fig. 4A shows a 3D iso-surface of a hexagonal unit. There are two twisted three-pronged triangles in the centre of the hexagonal unit, which actually lie in different triangular planes of C. Summing the electron density in the three-pronged triangles shows that each triangle contains approximately one electron. According to the Bader charge analysis (*SI Appendix*, Table. S4), on average, each Mn atom loses 0.6 $e^-$ and each C atom gains about 1.3 $e^-$. Therefore, the three C atoms in the triangular plane of C have exactly one spare electron between them. This confirms the previous observations about C atoms in Fig. 4A. Furthermore, there is some difference in the electron loss of each Mn atom, in that the Mn atoms in the medial triangular units in Fig. 1D lose more electrons than those in the lateral hexagonal units in Fig. 1C.

From this, it can be concluded that the C atoms within each three-pronged triangle each share about 1/3 of an electron, forming an electron cloud with 1 $e^-$, similar to electronic compounds (Fig. 4B). This electron is randomly acquired from one of the nearby Mn triangles, which leads to unequal electron loss and spin frustration of the medial triangular Mn atoms. The p-d orbital hybridisation introduces a sphere-like electron cloud around each C atom. In addition, the electron localisation function of the electrons around each C atom presents mirror symmetry (Fig. 4C), which resembles the polarisation of the C atoms and spin exchange. This polarisation is enhanced by cooling, which explains the sudden increase in DC susceptibility below 25 K. There is no relaxation time for this polarisation; therefore, the coherence length of the C atom is zero, which also explains why the imaginary part of the AC susceptibility is always zero and why the DC susceptibility increases with cooling.

Based on the above discussion, we hypothesise that $Mn_7C_3$ is a spin glass with good conductivity, and its complexity is mainly due to the triangular correlation between Mn atoms, while the C atoms act as interval atoms and electron `snatchers'. This unprecedented spin-glass system combines the characteristics of Ruderman-Kittel-Kasuya-Yosida (RKKY) and super-exchange interactions. This explains why the value of $z_v$ is closest to that of classical transition-metal solute spin glasses, while the derivation of the fitting line is closest to that of dilute magnetic semiconductor spin glasses (Fig. 2C).

To further verify this hypothesis, we calculated the spin density of $Mn_7C_3$. In Fig. 4D, the red and blue regions represent up- and down-spins, respectively. The different types of lines indicate different interactions and zones which are key to understanding the distinct magnetic behaviour of $Mn_7C_3$. The white dotted line outlines one of the antiferromagnetic exchange-correlation interaction paths, and the equilateral triangular zones marked with black dot-dashed lines are the medial triangular structures. The area in the black rectangle enclosing one side of the lateral honeycomb hexagonal structure is enlarged in the right panel of Fig. 4D to specify the role of magnetic exchange in the zig-zagged Mn chains and vertex atoms.

To demonstrate this more clearly, a schematic model of the atoms and spin directions is shown in Fig. 4E. The atoms and spin directions here have a one-to-one correspondence with those marked by the black dot-dashed and white dotted lines in Fig. 4D. Similar to the atoms in the medial triangular units, the atoms in the rectangular zone are also combined by antiferromagnetic coupling into an equilateral triangle. Notably, every Mn triangular unit is magnetically frustrated, regardless of whether it is positioned within the hexagon or at the edge of the honeycomb unit. Therefore, micro-spin competition occurs throughout the $Mn_7C_3$ crystal, and the magnetic order of a third atom cannot satisfy the antiferromagnetic coupling of the other two atoms in any triangle. The energy of any triangle unit must be equal, regardless of whether the magnetic order of the third atom is up or down[25].

The final magnetic state of the third atom is determined by `external disturbance', such as the electron snatching behaviour of the triangular C atoms. In addition, the spin polarisation of the C atoms acts as a messenger state (Fig. 4C), as do the Mn atoms comprising the hexagonal cage. These messenger states transmit the spin exchange between the honeycombs and triangular units. Furthermore, the C atoms connect the whole crystal, acting as a bridge that form a complex 3D frustrated spin network. All of the triangular units meet the Ising frustration model[16]. Further, all the triangular units could be at completely different magnetic orders in this system, and the degeneracy would still be natural. Small disturbances to the spin configurations would affect the competition among the magnetic atoms and result in an irregular distribution of the entire magnetic status of the bulk material; therefore, the spin-glass state is inevitable.

## Discussion

In summary, we synthesised a high-quality bulk crystalline *Pnma*-type $Mn_7C_3$ compound by a high-temperature and high-pressure method. The $Mn_7C_3$ structure consists of infinite triangular Mn units with C atoms between the medial triangular Mn units. The microgeometry of the structure bestows it with spin-glass behaviour with a $T_f$ of 37.4 K. The frustration within each triangular Mn unit, which is caused by the random capture of shared electrons by the triangle C atoms, drives competition of the antiferromagnetic interaction between the three equidistant Mn atoms. Meanwhile, owing to the influence of p-d orbital hybridisation, the spin polarisation of C atoms bridges the exchange-correlation from different magnetic triangular units. This polarisation is gradually enhanced as the temperature decreases, which leads to an increase of the magnetic susceptibility at 20 K and lower. The competition and frustration between different magnetic triangular units bestow the $Mn_7C_3$ crystal with spin-glass behaviour and form its complex frustrated spin network. The triangular units are a natural geometric characteristic of $Mn_7C_3$; therefore, in addition to being as robust as other spin glasses, $Mn_7C_3$ is also a specific self-induced spin glass that does not rely on atomic disorder or displacement. This kind of triangular unit contains many degenerate states that could be useful in the fields of data storage and artificial intelligence. Finally, as a representative structure comprising infinite natural triangular units, the spin-glass behaviour of $Mn_7C_3$ shows that other magnetic materials with similar crystal structures and antiferromagnetic interactions could also be spin glasses. These results expand our understanding of complex spin-glass systems.

Supporting Information for

**Self-induced crystalline fluctuation spin-glass state in $Mn_7C_3$ binary compounds**

Zekun Yu[1,2], Chao Zhou[1], Kuo Bao[1*], Xiaofeng Wang[3], Zhaoqing Wang[1], Jinming Zhu[1], Enxuan Li[1], Andong Yao[1], Yuhan Meng[1], Yufei Ge[1], Xingbin Zhao[4], Shuailing Ma[4], Pinwen Zhu[1], Qiang Tao[1], Tian Cui[1,4*]

[1]State Key Laboratory of Superhard Materials, College of Physics, Jilin University, Changchun, 130012, China

[2]Center for Optics Research and Engineering, Shandong University, Qingdao, 266237, China

[3]State Key Laboratory of Inorganic Synthesis and Preparative Chemistry, College of Chemistry, Jilin University, Chang chun 130012, China

[4]Institute of High Pressure Physics, School of Physical Science and Technology, Ningbo University, Ningbo 315211, China

*Corresponding authors: baokuo@jlu.edu.cn, and cuitian@nbu.edu.cn

**Table S1. EDS data for Mn and C**

| Element | Type | Concentration | K-factor | wt% | wt% Sigma | Atomic percent |
|---|---|---|---|---|---|---|
| Mn | K linellae | 19.82 | 0.19822 | 93.79 | 0.1 | 76.75% |
| C | K linellae | 0.61 | 0.00611 | 6.21 | 0.1 | 23.25% |

**Table S2. Reference data from JCPDS No. 75-1498 and crystal data from Rietveld refinement**

| Identification code | | | cal-$Mn_7C_3$ | exp-$Mn_7C_3$ |
|---|---|---|---|---|
| Molecular weight (u) | | | 420.613 | 420.613 |
| Radiation | | | Mo | Mo |
| Wavelength (Å) | | | 0.7093 | 0.7093 |
| Crystal system | | | Orthorhombic | Orthorhombic |
| Space group | | | $Pnma(62)$ | $Pnma(62)$ |
| Unit cell dimensions (Å) | | | $a = 4.546$ | $a = 4.520(3)$ |
| | | | $b = 6.959$ | $b = 6.913(7)$ |
| | | | $c = 11.976$ | $c = 11.919(8)$ |
| Cell volume (Å$^3$) | | | 378.87 | 372.52 |
| Calculated density (g·cm$^{-3}$) | | | 7.372 | 7.499 |
| Range (°) | | | 10-45 | 10-45 |
| Residuals | | | None | $R_p$ = 4.11% |
| | | | None | $R_{wp}$ = 5.90% |
| | | | None | $\chi^2 = 0.038$ |
| **Atomic position** | **Wyckoff** | **Occupation** | **cal - (x, y, z)** | **exp - (x, y, z)** |
| Mn1 | 8$d$ | 1 | (0.25, 0.07, 0.02) | (0.26, 0.05, 0.02) |
| Mn2 | 8$d$ | 1 | (0.00, 0.57, 0.19) | (0.06, 0.80, 0.08) |
| Mn3 | 4$c$ | 1 | (0.24, 0.25, 0.20) | (0.28, 0.25, 0.18) |
| Mn4 | 4$c$ | 1 | (0.25, 0.25, 0.42) | (0.28, 0.25, 0.42) |
| Mn5 | 4$c$ | 1 | (0.00, 0.25, 0.63) | (0.10, 0.25, 0.64) |
| C1 | 8$d$ | 1 | (0.11, 0.03, 0.35) | (0.04, 0.04, 0.35) |
| C2 | 4$c$ | 1 | (0.38, 0.25, 0.57) | (0.46, 0.25, 0.56) |

A

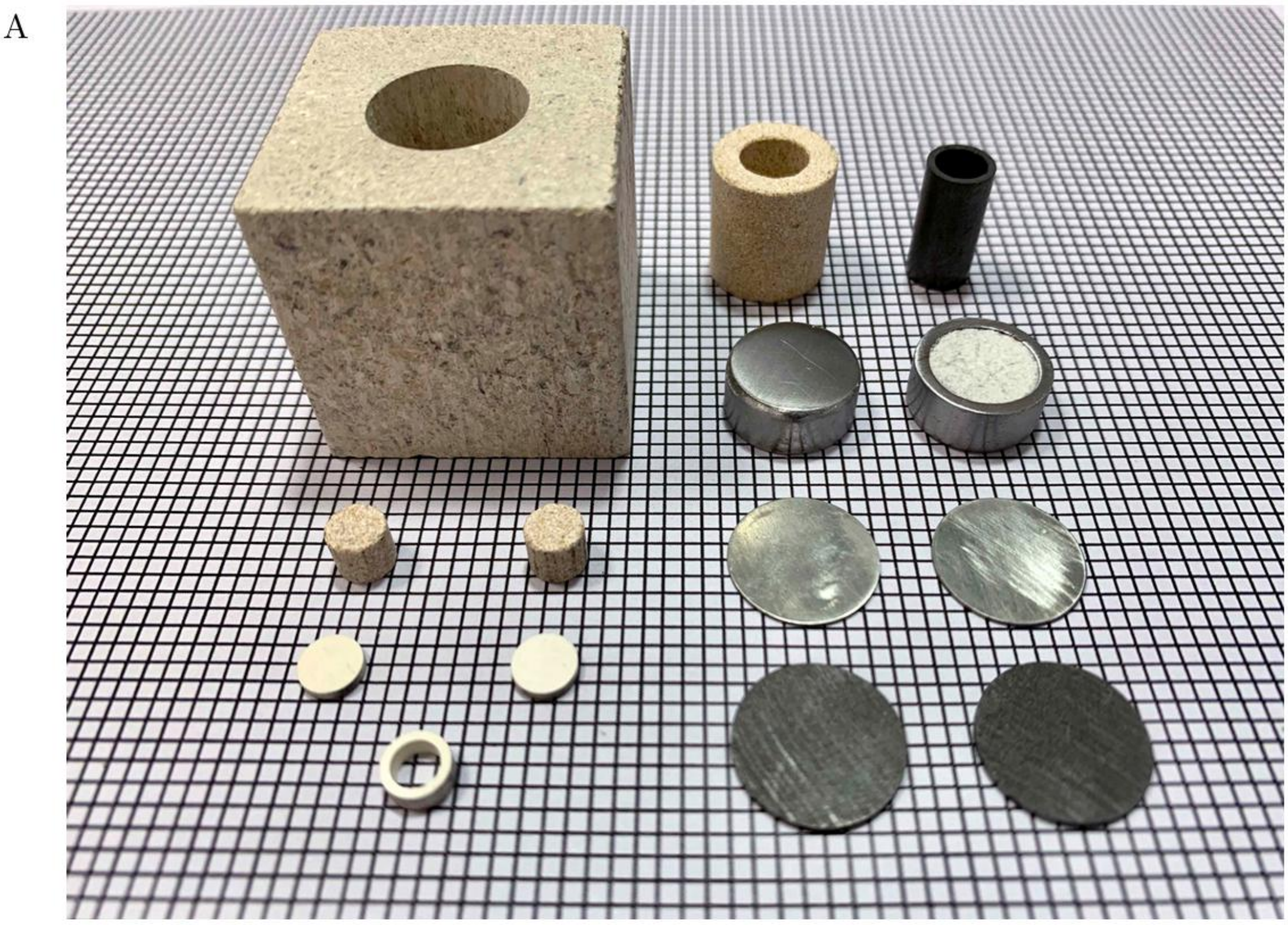

B

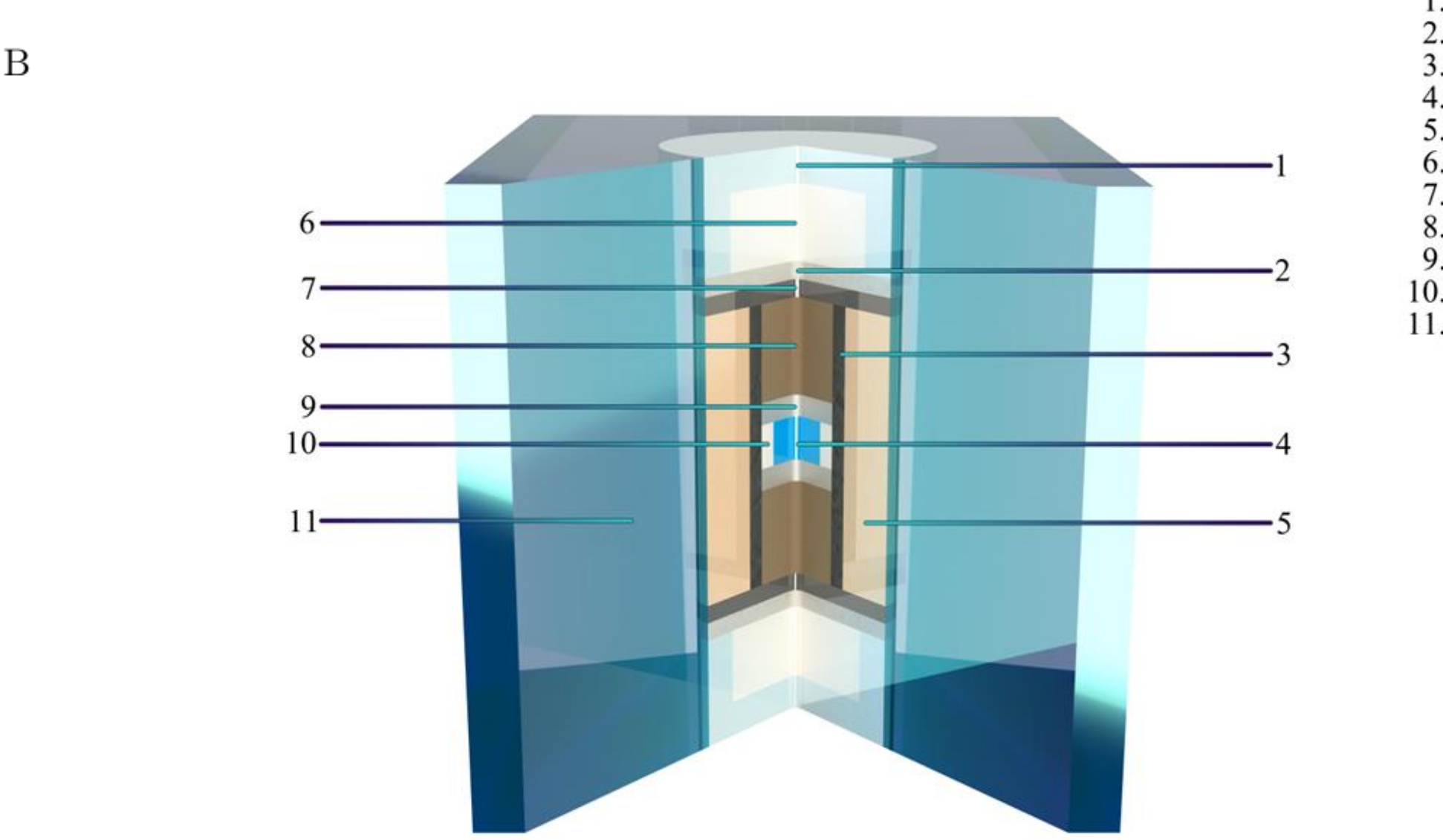

1. Steel gasket
2. Mo chip
3. Graphite tube
4. Sample
5. MgO tube
6. MgO bulk
7. Graphite chip
8. BN cylinder
9. BN chip
10. BN tube
11. Pyrophyllite

**Fig. S1.** Assembly of high-temperature, high-pressure synthesis unit. (A) Actual components of the synthesis cell (without sample). The mesh grid size is 2 mm. (B) Schematic of the assembled synthesis cell. The part names are shown in the legend.

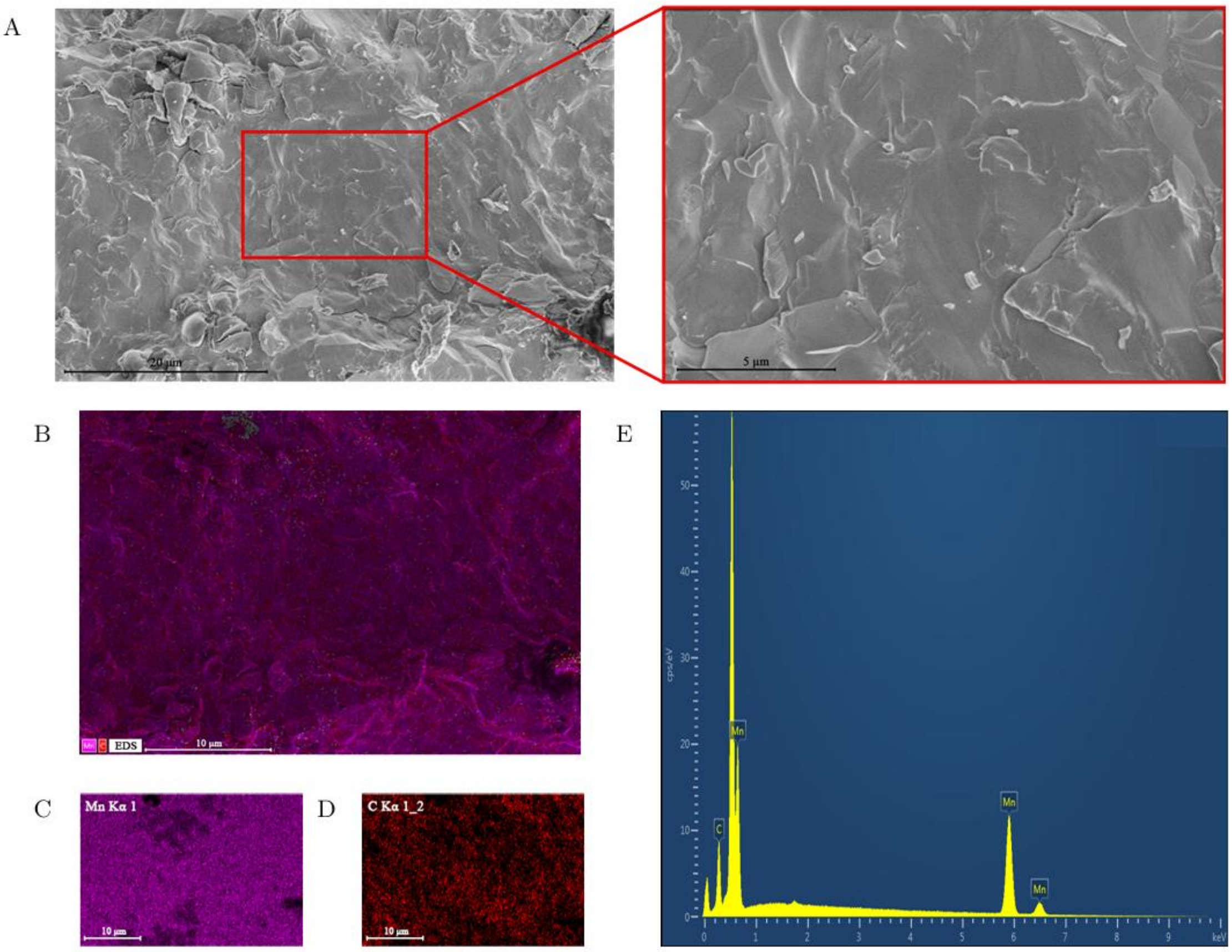

**Fig. S2.** Morphology and elemental distribution of Mn7C3 sample. (A) SEM images showing the surface topography at low (left) and high magnifications (right). (B) EDS elemental mapping image of Mn (purple) and C (red) in area (A). (C) and (D) are the individual Mn and C distributions. (E) Energy dispersive spectrum of sample.

**Table S3. Initial spin configurations of random, parallel, and antiparallel states**

| Spin state | Initial configuration |
|---|---|
| Random | $\lvert\psi\rangle_{\text{random}} = \lvert\uparrow\downarrow\downarrow\downarrow\uparrow\downarrow\uparrow\uparrow\downarrow\downarrow\downarrow\uparrow\downarrow\uparrow\rangle\lvert\downarrow\uparrow\uparrow\uparrow\downarrow\uparrow\downarrow\downarrow\uparrow\uparrow\uparrow\downarrow\uparrow\downarrow\rangle$ |
| Parallel | $\lvert\psi\rangle_{\text{parallel}} = \lvert\uparrow\downarrow\uparrow\downarrow\uparrow\downarrow\uparrow\downarrow\uparrow\downarrow\uparrow\downarrow\uparrow\downarrow\rangle\lvert\uparrow\downarrow\uparrow\downarrow\uparrow\downarrow\uparrow\downarrow\uparrow\downarrow\uparrow\downarrow\uparrow\downarrow\rangle$ |
| Antiparallel | $\lvert\psi\rangle_{\text{antiparallel}} = \lvert\uparrow\uparrow\uparrow\uparrow\uparrow\uparrow\uparrow\uparrow\uparrow\uparrow\uparrow\uparrow\uparrow\uparrow\rangle\lvert\downarrow\downarrow\downarrow\downarrow\downarrow\downarrow\downarrow\downarrow\downarrow\downarrow\downarrow\downarrow\downarrow\downarrow\rangle$ |

A

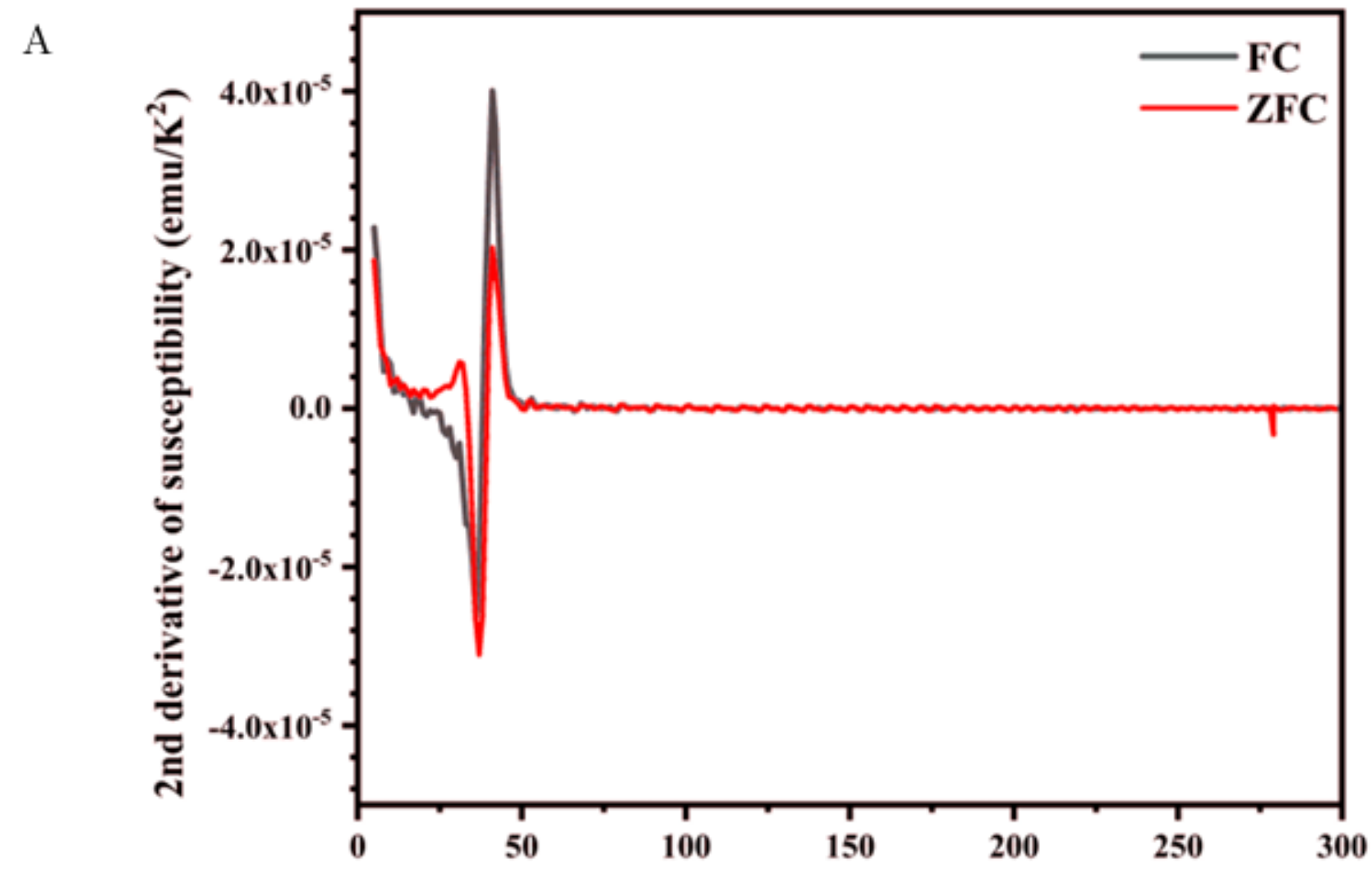

B

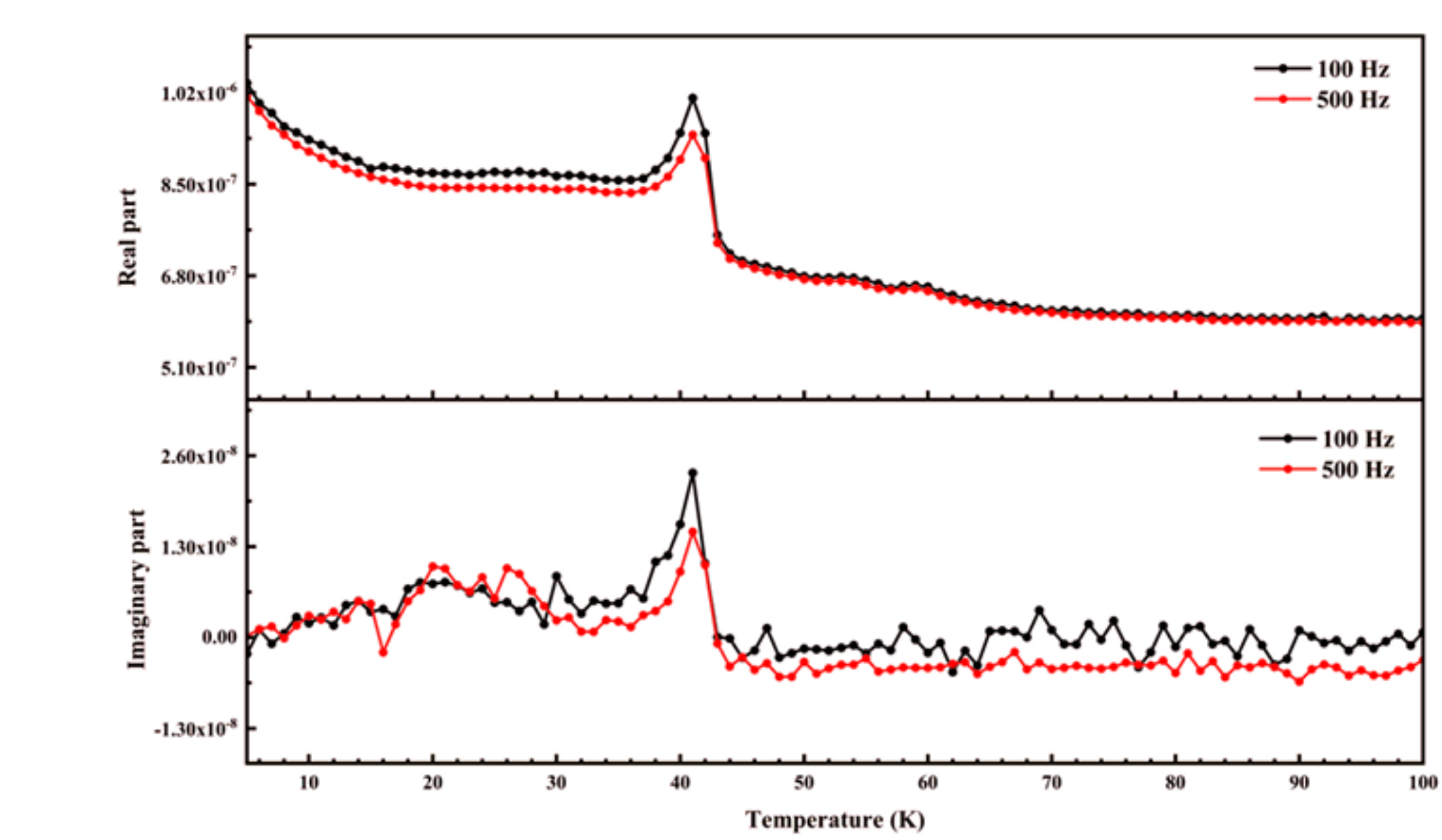

**Fig. S3.** Second derivative of DC magnetic susceptibility and real and imaginary parts of AC magnetic susceptibility. (A) Second derivative of DC magnetic susceptibility for samples with FC (black line) and ZFC (red line). (B) Temperature dependence of the real (upper) and imaginary (lower) parts of AC magnetic susceptibility at field frequencies of 100 Hz (black line) and 500 Hz (red line).

**Table S4. Bader charge results of random, parallel and antiparallel state**

| Element | | | Charge ($e^-$) | | |
|---|---|---|---|---|---|
| | | | Random | Parallel | Antiparallel |
| Mn | Lateral | 1 | 0.565456 | 0.569575 | 0.538643 |
| | | 2 | 0.552946 | 0.569570 | 0.538137 |
| | | 3 | 0.537927 | 0.569702 | 0.567662 |
| | | 4 | 0.545104 | 0.569277 | 0.567051 |
| | | 5 | 0.543565 | 0.538105 | 0.566197 |
| | | 6 | 0.538203 | 0.538523 | 0.566967 |
| | | 7 | 0.552822 | 0.538446 | 0.539764 |
| | | 8 | 0.565058 | 0.538029 | 0.539887 |
| | | 9 | 0.458091 | 0.471050 | 0.477116 |
| | | 10 | 0.477490 | 0.471050 | 0.477387 |
| | | 11 | 0.459055 | 0.471975 | 0.476454 |
| | | 12 | 0.476694 | 0.471975 | 0.475833 |
| | | 13 | 0.559046 | 0.532358 | 0.533616 |
| | | 14 | 0.530471 | 0.532358 | 0.533654 |
| | | 15 | 0.557550 | 0.532697 | 0.533805 |
| | | 16 | 0.529219 | 0.532697 | 0.534328 |
| | Medial | 1 | 0.589624 | 0.604101 | 0.566569 |
| | | 2 | 0.596113 | 0.604101 | 0.566331 |
| | | 3 | 0.575675 | 0.604821 | 0.599180 |
| | | 4 | 0.575926 | 0.604821 | 0.598807 |
| | | 5 | 0.575629 | 0.605348 | 0.599229 |
| | | 6 | 0.575658 | 0.605348 | 0.599167 |
| | | 7 | 0.596103 | 0.604036 | 0.566402 |
| | | 8 | 0.590901 | 0.604036 | 0.566566 |
| | | 9 | 0.590807 | 0.588513 | 0.584772 |
| | | 10 | 0.581498 | 0.588513 | 0.584298 |
| | | 11 | 0.590934 | 0.588798 | 0.585717 |
| | | 12 | 0.581999 | 0.588798 | 0.584521 |
| C | | 1 | 1.290085 | 1.290152 | 1.280059 |
| | | 2 | 1.290904 | 1.290152 | 1.279963 |
| | | 3 | 1.276419 | 1.289135 | 1.289496 |
| | | 4 | 1.278216 | 1.289135 | 1.290295 |
| | | 5 | 1.278946 | 1.292480 | 1.289168 |
| | | 6 | 1.275589 | 1.292480 | 1.288817 |
| | | 7 | 1.291537 | 1.291111 | 1.280335 |
| | | 8 | 1.289693 | 1.291111 | 1.279830 |
| | | 9 | 1.307953 | 1.327716 | 1.297765 |
| | | 10 | 1.291967 | 1.327716 | 1.298040 |
| | | 11 | 1.306778 | 1.328717 | 1.297064 |
| | | 12 | 1.291479 | 1.328717 | 1.297235 |